\documentclass[useAMS,usenatbib]{mn2e}
\usepackage[dvips]{graphicx,color}
\usepackage{txfonts}
\usepackage{amssymb}
\usepackage{indentfirst}
\usepackage{epsfig}

\newcommand{\CaXX}{\mbox{{Ca\,{\sevensize XX}}}}

\newcommand{\FeXXV}{\mbox{{Fe\,{\sevensize XXV}}}}

\newcommand{\Lya}{\ensuremath{\hbox{Ly}\alpha~}}
\newcommand{\Ka}{\ensuremath{\hbox{K}\alpha~}}
\newcommand{\Kb}{\ensuremath{\hbox{K}\beta~}}

\def\cha{{\it Chandra }}

\def\nus{{\it NuSTAR }}

\begin{document}
\title[X-ray fluorescent lines of M51]{X-ray fluorescent lines from the Compton-thick AGN in M51} 

\author[W. Xu et al.]{Weiwei Xu, Zhu Liu, Lijun Gou, and  Jiren Liu \\
National Astronomical Observatories, 20A Datun Road, Beijing 100012, China; 
{weiweixu@nao.cas.cn, jirenliu@nao.cas.cn}\\
}
\date{}

\maketitle

\begin{abstract}

The cold disk/torus gas surrounding active galactic nuclei (AGN) emits fluorescent lines 
when irradiated by hard X-ray photons. The fluorescent lines of elements 
other than Fe and Ni are rarely detected due to their relative faintness.
We report the detection of \Ka lines of neutral Si, S, Ar, Ca, Cr, 
and Mn, along with the prominent Fe K$\alpha$, Fe K$\beta$, and Ni \Ka lines,
from the deep \cha observation of the low-luminosity Compton-thick AGN in M51.
The Si \Ka line at 1.74 keV is detected at $\sim3\sigma$, the other fluorescent
lines have a significance between 2 and 2.5 $\sigma$, while the Cr line has a
significance of $\sim1.5\sigma$.
These faint fluorescent lines are made observable due to the heavy obscuration
of the intrinsic spectrum of M51, which is revealed by \nus observation above 10 keV.
The hard X-ray continuum of M51 from \cha and \nus can be fitted with a 
power-law spectrum with an index of 1.8, reprocessed by a torus with an equatorial column 
density of $N_{\rm H}\sim7\times10^{24}$ cm$^{-2}$ and an inclination angle of $74$ degrees.
This confirms the Compton-thick nature of the nucleus of M51.
The relative element abundances inferred from the fluxes of the fluorescent lines are similar 
to their solar values, except for Mn, which is about 10 times overabundant.
It indicates that Mn is likely enhanced by the nuclear spallation of Fe.

\end{abstract}


\section{Introduction}

The majority of nearby active galactic nuclei (AGN) are
obscured by large amounts of cold gas and dust \citep[e.g.][]{Com04}.
For Compton thick AGN with absorption column densities larger than
$1.5\times10^{24}$ cm$^{-2}$ (the inverse of the Thomson cross-section), 
the continuum below 10 keV is heavily suppressed, and
Compton scattering of high energy photons produces a spectral bump around $20-30$ keV. 
The cold gas emits fluorescent lines when irradiated by the 
central AGN. The most prominent line is the Fe K$\alpha$ line at 6.4 keV, due to its 
high abundance and high fluorescence yield. The Fe 6.4 keV line is 
found to be a ubiquitous feature of AGN \citep[e.g.][and reference therein]{Shu10}.
Since the continuum is heavily suppressed for Compton-thick AGN, 
their Fe 6.4 keV line will be more noticeable.
Indeed, the Fe 6.4 keV lines with equivalent width (EW) as large as a few keV have been 
reported \citep[e.g.][]{Lev02}.

For those AGN with noticeable Fe K$\alpha$ lines, the fluorescent lines 
from elements other than Fe should also be detectable. 
The fluorescent lines of other elements 
are generally thought to be observationally irrelevant, due to their small yields.
Nevertheless, the yields of other elements are not as small as unobservable. For example, 
the yield of neutral Si K$\alpha$ is 0.042, larger than 10\% of that of the neutral Fe K$\alpha$.
The fluxes of the fluorescent lines are determined by the chemical abundances, the distribution 
of the obscuring material, as well as the illuminating intrinsic spectrum of AGN. 
If detected, the non-Fe fluorescent lines can be used to measure the abundances of the cold gas.
A possible example is the {\it ASCA} observation of NGC 6552 that shows K$\alpha$ lines
of seven neutral species, which were used to constrain the abundances of the 
cold gas \citep{Fuk94,Rey94}.

In this letter we report the detection of fluorescent lines of 
neutral Si, S, Ar, Ca, Cr, and Mn, along with the prominent Fe K$\alpha$, Fe K$\beta$, 
and Ni \Ka lines, from the deep \cha spectrum of the nucleus of M51 (NGC 5194).
Known as the Whirlpool galaxy, M51 is classified as a LINER/Seyfert II galaxy 
\citep[e.g.][]{Sta82,Ho97}. 
Its face-on inclination and close distance \citep[7.1 Mpc,][]{TV06}
allow detailed studies of the nuclear activity \citep[see e.g.][and references therein]{Liu15}.
While the X-ray luminosity of the nucleus of M51 in $2-10$ keV band is as low 
as $\sim10^{39}$ erg s$^{-1}$, the EW of the Fe 6.4 keV line is $\sim4$ keV, 
the largest one ever reported \citep[e.g.][]{TW01,Lev02}. 
Such a large EW implies a column density larger than $3\times10^{24}$ cm$^{-2}$ as
shown by Monte-Carlo simulations of Compton-thick torus \citep{Ike09,MY09}.

A bright hard X-ray excess above 10 keV had been reported from the 
{\it BeppoSAX} observation of M51 \citep{Fuk01}, 
and an absorption column density $\sim10^{24}$ cm$^{-2}$ was inferred.
The lack of spatial information of {\it BeppoSAX}, however, makes the inference uncertain.
Indeed, as shown in \S 2, the \nus observation of M51 detects at least three sources 
within its field of view (FOV).
We include the \nus data of the nucleus of M51 to constrain the continuum above 10 keV.

\section{Observational data }

\begin{figure*}
\includegraphics[height=2.5in]{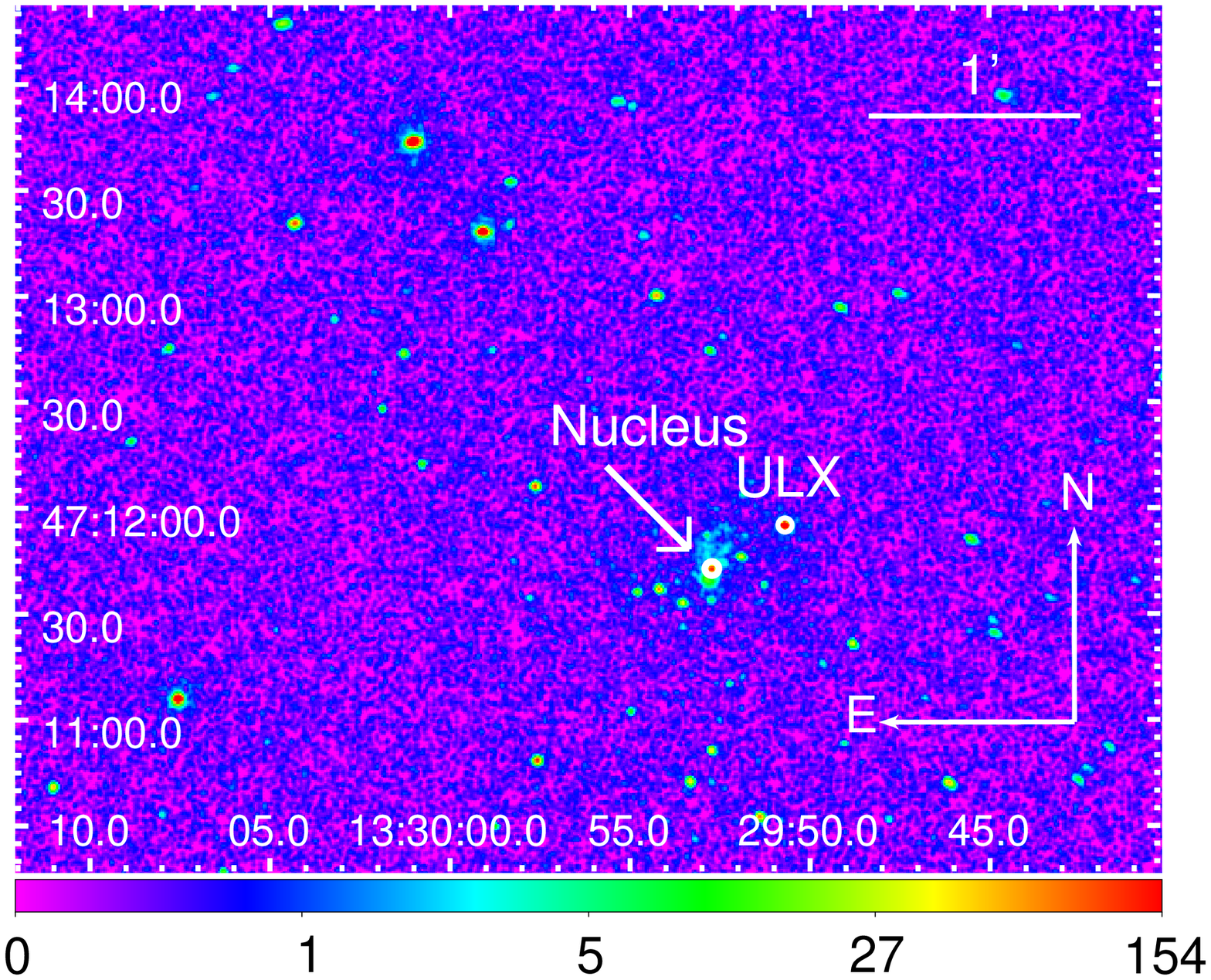}
\includegraphics[height=2.5in]{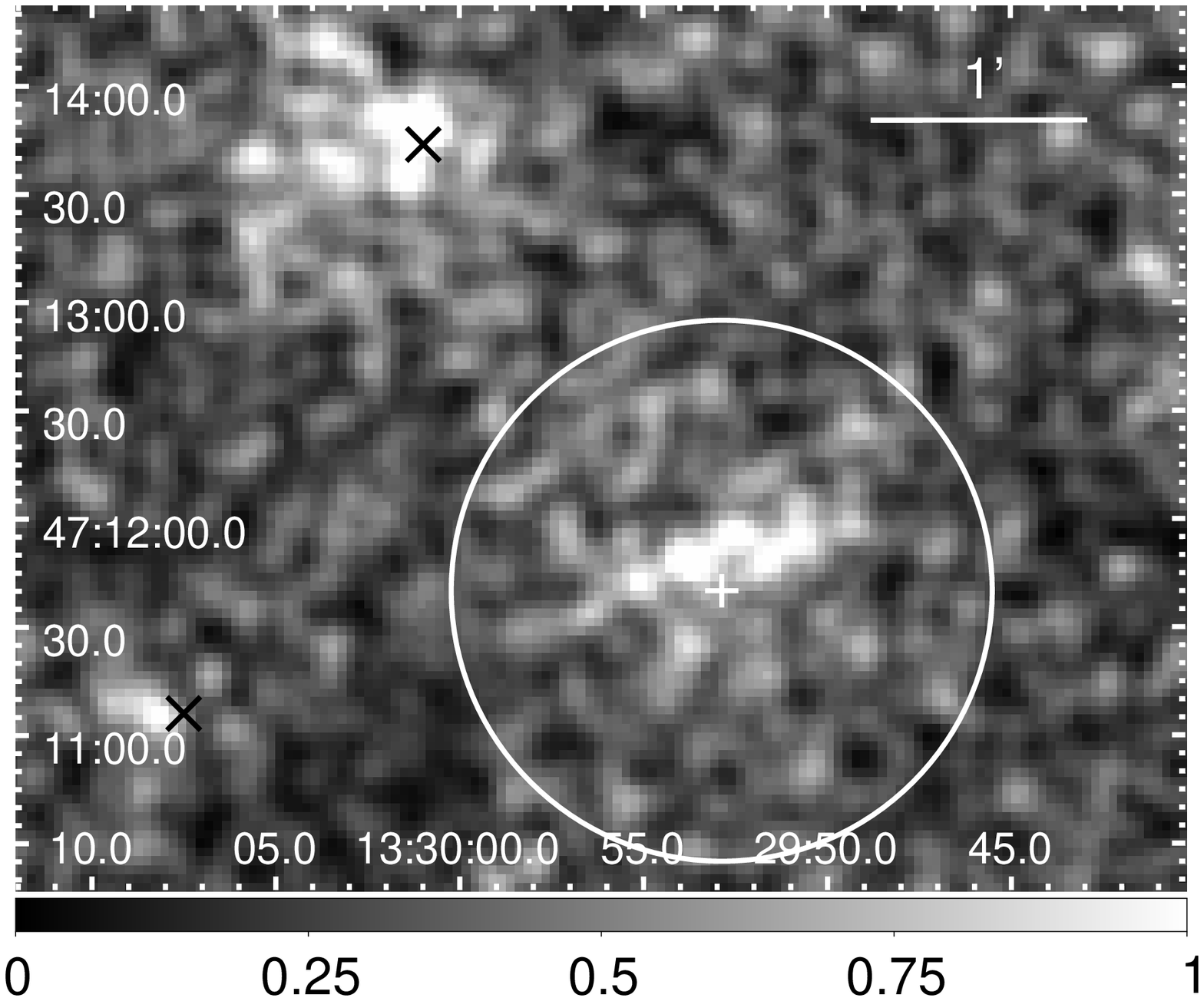}
\caption{
	Left: \cha image of M51 within $2-7.5$ keV band merged from all twelve datasets.
	The nuclear emission is dominated by a ULX 25$''$ away from the nucleus.
	Right: \nus image of M51 combining both FPMA and FPMB data. The white plus sign
	marks the nucleus of M51, and the circle with a radius of 75$''$
	indicates the spectral extraction region. The two black crosses indicate the 
	other two sources detected
	by {\it NuSTAR}. The \cha and \nus images are smoothed 
	with a Gaussian of 2 ($1''$) and 3 ($7.4''$) pixels, respectively. 
}
\end{figure*}

The half-arcsec spatial resolution of \cha makes it possible to resolve the point sources 
in the nuclear region of M51. This helps to reduce the contamination of other sources 
to the nucleus of M51. 
We use twelve \cha archival observations with ObsID numbers of 13812, 13813, 13814,
13815, 13816, 15496, 15553, 1622, 354, 3932, 12562, and 12668, all of which were taken with
ACIS-S. 
The datasets are analyzed with {\it CIAO} (version 4.6) following standard procedures.
After removing the flare periods, the total effective exposure time is about 800 ks,
and the total count number is about 12,000 for the nucleus.
All datasets are aligned to each other using the {\it CIAO} tools 
of {\it wcs\_match} and {\it wcs\_update}.
As an illustration, the counts image of M51 within $2-7.5$ keV band merged
from all datasets is plotted in the left panel of Figure 1. 
It is dominated by an off-nuclear ULX.
While the nucleus is relatively faint within 
$2-7.5$ keV band, its striking 6.4 keV line (as shown
in Figure 2) indicates the hidden AGN. The extended emission can  
be clearly seen in Figure 1, which has been studied previously \citep[e.g.][]{OW09}.

M51 was observed with \nus for a short exposure of 18.5 ks on Oct. 2012.
The data is reduced using NUSTARDAS software and the effective exposure is 
about 15 ks after cleaning. In the right panel of Figure 1, we show the 
\nus image of M51 combining the two focal plane modules of FPMA and FPMB.
Besides the nucleus, there are two other sources detected by {\it NuSTAR}
(indicated as black crosses in Figure 1), which correspond to sources 69 (upper one) 
and 82 (lower-left one) as in \citet{TW04}.

\section{Results}

\begin{figure}
\hspace*{-0.5cm}
\includegraphics[height=2.2in]{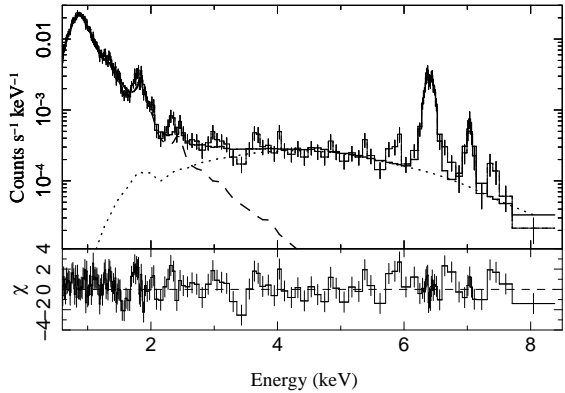}
\caption{
	\cha spectrum of the nucleus of M51 merged from all twelve datasets. 
	The fitted model of Vapec (dashed line) plus a powerlaw (dotted line) and two Gaussians is 
	over-plotted. Many residual peaks are around the energies of \Ka lines of neutral species. 
}
\end{figure}

\begin{table}
\caption{Model of MYTorus+Vapec+Gaussians fitted to the spectra of the nucleus of M51}
\begin{tabular}{lcccc}
\hline
\hline
MYTorus+Vapec & Value \\
\hline
Intrinsic powerlaw index, $\Gamma$ & $1.8\pm0.3$ \\
Intrinsic powerlaw normalization  & $0.002\pm0.0015$ \\
MYTorus equatorial column density ($10^{24}$ cm$^{-2}$)$^a$ & $7.0\pm3.0$ \\
MYTorus inclination angle (degrees) & $74\pm8$ \\
Absorption column density of Vapec ($10^{22}$ cm$^{-2}$) & $0.05\pm0.04$ \\
Vapec temperature (keV) & $0.74\pm0.03$ \\
Vapec normalization ($10^{-5}$) & $7.1\pm1.4$\\
Vapec O & $0.3\pm0.3$\\
Vapec Ne & $0.5\pm0.3$\\
Vapec Mg, Al & $0.2\pm0.1$\\
Vapec Si, S, Ar, Ca & $0.3\pm0.1$\\
Vapec Fe, Ni & $0.2\pm0.1$ \\
$\chi^2/d.o.f.$ & $169.3/155\sim1.1$ \\
\hline
\end{tabular}
\vspace*{-0.2cm}
 \tabcolsep 3.02pt
\begin{tabular}{lccccccc}
      \hline
		Line & $\rm Energy (keV)$ &  Flux & Y$_x$ & $\rm Z_{x}/Z_{Fe}$ &$\Delta\chi^2$$^b$ & Sig.$^c$   \\
      \hline
		Si K$\alpha$  &   1.74 &       $1.9\pm0.8$     & 0.042  &$0.8\pm0.3$ &16.0&99.6\%     \\
S  K$\alpha$    & 2.31   &       $1.3\pm0.7$    &0.078&$0.6\pm0.3$         &8.3 &97.4\%\\
 Ar K$\alpha$   &    2.96 &      $0.8\pm0.5$    &0.112 &$0.9\pm0.6$        &6.2 &97.0\%\\
 Ca K$\alpha$   & 3.69    &    $0.9\pm0.5$      &0.124 &$1.3\pm0.7$        &8.7 &97.5\%\\
 Cr K$\alpha$   &     5.41&     $0.7\pm0.6$     &0.25 &$2.2\pm1.9$         &2.4 &85.8\% \\
 Mn K$\alpha$   &    5.89   &     $1.8\pm0.9$   &0.278 &$9.8\pm4.9$        &8.8 &97.7\%\\
Fe K$\alpha$    &  $6.40\pm0.01$  &$38.0\pm3.3$&0.304  & 1                 &--&--\\
 Ni K$\alpha$   &    $7.42\pm0.04$ &$4.4\pm2.2$ &0.37 &$2.2\pm1.1$         &10.2&98.3\%\\
Fe K$\beta$     &   $7.03\pm0.02$&$7.8\pm2.2$    &-- &  --                 &--&-- \\
\FeXXV\ K$\alpha$    &  6.7     &     $1.8\pm1.3$ &-- &--                     &5.3&96.5\%\\
        \hline

\end{tabular}
\begin{description}
\begin{footnotesize}
\item
  Note: The abundances of Vapec are relative to their solar values. All the line energies 
  are fixed at expected values, except for Fe K$\alpha$, Fe K$\beta$, and
  Ni K$\alpha$, which are fitted. The line flux is in units of $10^{-7}$ photons cm$^{-2}$ s$^{-1}$.
  The yield Y$_x$ is taken from \citet{KM93}.
  The errors quoted are for 90\% confidence level. $^a$The upper limit of the equatorial column density
  is beyond the valid $N_{\rm H}$ value of MYTorus model of $10^{25}$ cm$^{-2}$. $^b$$\Delta\chi^2$
is the improvement of $\chi^2$ by adding the line. $^c$The significance is estimated based
on Monte-Carlo simulations.
\end{footnotesize}
\end{description}
\end{table}

The spectrum of the nucleus of M51 is extracted from a 2$''$ radius for each \cha
observation with a background spectrum extracted from a source-free region.
We compared the spectra of different observations
and found no apparent variability. Thus we combine them using the ISIS function 
{\it combine\_datasets}, for which the model is calculated for each dataset,
and the datasets and models are summed separately before computing the $\chi^2$
\citep{ISIS}. 
The combined spectrum is plotted in Figure 2, which is binned to a minimum 
signal-to-noise (S/N) ratio of 4.5 below 6.8 keV
and 3 above that to enhance the visibility of emission lines of low counts.
We see that the spectrum has a significant soft component below 2 keV, which is 
likely due to the shock-heated gas by the radio jet \citep{TW01}.
While above 2 keV, the most prominent feature are the emission lines at 6.4 and 7 keV, which are
the K$\alpha$ and K$\beta$ lines of neutral-like Fe. 
In addition, there are also many emission spikes between 2 and 6 keV.

We first try to fit the \cha spectrum of M51 with a thermal model \citep[Vapec,][]{Fos12},
representing the soft component,
plus an absorbed powerlaw model for the hard component. We include two Gaussians to represent 
the Fe K$\alpha$ and K$\beta$ lines. 
The fitted result is plotted in Figure 2. 
The fitted temperature of the thermal component is about 0.7 keV, a little higher 
than the value reported by \citet{TW01}. The fitted powerlaw index is about -0.3,
implying a heavily reprocessed continuum.
In the residual panel of Figure 2, there 
are many remaining spikes, such as the ones around 1.7 keV, 2.3 keV, 5.9 keV, and 7.5 keV. 
We find they are exactly corresponding to the K$\alpha$ emission lines of neutral Si, S, Mn, and Ni.
Therefore, they are most likely due to the fluorescent lines of neutral species as the Fe 6.4 keV line. 

\begin{figure*}
\hspace*{-0.8cm}
\includegraphics[height=2.8in]{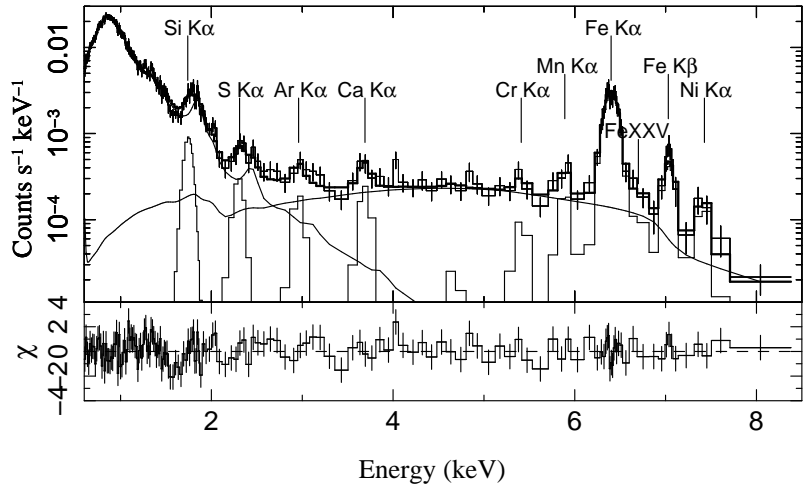}
\caption{
The model of Vapec+MYTorus+Gaussians fitted to the \cha spectrum of the nucleus of M51.
Each component is overlaid individually. The residuals around the added \Ka lines are 
reduced significantly. The model line feature around 4.6 keV is the instrumental Si escape 
peak, which follows the Fe 6.4 keV line and is about 1\% of the intensity of the Fe line \citep{Gri09}.
}
\end{figure*}

To measure the observed line fluxes, it is better to use a scattered continuum rather than a powerlaw.
For this purpose, we adopt the MYTorus model \citep{MY09, Yaq12}, which is based on 
Monte-Carlo simulations of a toroidal geometry.
The MYTorus model is specified by the incident powerlaw continuum, the equatorial column density,
and the inclination angle. It provides separable components of transmitted
continuum, scattered continuum, and the Fe K$\alpha$ and K$\beta$ fluorescent lines 
for a solar abundance \citep{AG89}.
Fluorescent lines from other elements are not included. 
Thus we replace the powerlaw model with the MYTorus continuum, which includes the 
transmitted and scattered components. We then add Gaussian lines 
centred at the energies of the K$\alpha$ lines of neutral elements, including
Si, S, Ar, Ca, Cr, Mn, and Ni. We also add a Gaussian line at 6.7 keV, where seems to be 
a faint \FeXXV\ \Ka line.

To further constrain the continuum model of MYTorus, we include the \nus spectrum of M51
into the fitting, which is extracted from a circle region of 75$''$ radius as indicated
in Figure 1. The background spectrum is extracted from a source-free region. The spectra of FPMA
and FPMB are combined using ISIS function {\it combine\_datasets}. 
Because \nus has a half-power diameter of $58''$, its spectrum
of the nucleus of M51 is contaminated by neighbouring sources. 
The dominate one is the ULX as indicated in the left panel of Figure 1, 
the \cha spectrum ($2.3-7$ keV) of which
can be well fitted with an absorbed powerlaw with a photon index of 1.85 and a column density
of $6.5\times10^{22}$ cm$^{-2}$. We also extract a spectrum from a circle region of 
60$''$ radius excluding the nucleus and the ULX, which
can be fitted with a powerlaw with an index of 2.4 without absorption.
The region of 60$''$ radius includes most of the extended emission and point sources that contaminate
the \nus spectrum. Therefore, besides the model for the nucleus of \cha data, 
we also assign these two powerlaw models with fixed parameters to the \nus data.
The instrument normalization between \cha and \nus is fixed.

\begin{figure}
\hspace{-0.0cm}
\includegraphics[height=2.2in]{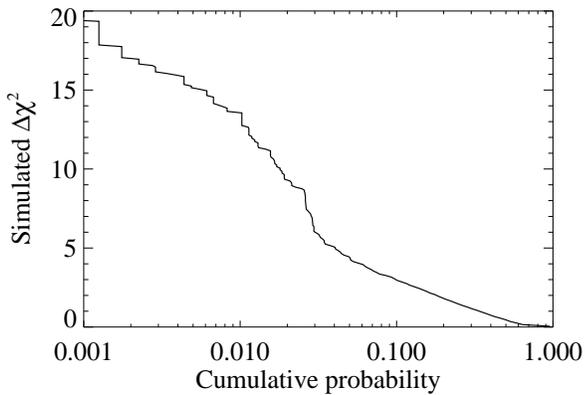}
\caption{
Distribution of the improvement of $\chi^2$ for the addition of an emission line 
for simulated fake spectra of a null hypothesis model including Vapec, a continuum, and
only Fe \Ka and \Kb lines. 
}
\end{figure}

The fitting results are listed in Table 1 and plotted in Figure 3 for the \cha data.
We see that the model provides a reasonable fit to the observed spectrum.
The improvement of $\chi^2$ with the addition of each line is also listed in Table 1.
There is a residual spike around 4.02 keV, which could be due to 
the fluorescent lines of Ca \Kb (4.01 keV), Sc \Ka (4.09 keV), and/or \CaXX\ \Lya line (4.11 keV).
If it is the Ca \Kb line, the measured Ca $K\alpha$/Ca \Kb ratio ($\sim1.5$) would be too small
compared with the expected value of 10 \citep{KM93}.
Since it is relatively weak ($\Delta\chi^2\sim4$ adding two parameters) and offset from
the expected fluorescent energies, we neglected it here.
The residuals below 2 keV are likely due to the simple model of one thermal component.
The fitted equatorial column density of the MYTorus model is $\sim7\times10^{24}$ cm$^{-2}$, close
to the up-limit of the valid column density of MYTorus. It confirms the Compton-thick nature
of the nucleus of M51. The fitted intrinsic powerlaw index is 1.8, typical of unobscured AGN.
The model luminosity within $2-10$ keV is $6.7\times10^{38}$ erg s$^{-1}$, while the intrinsic 
luminosity corrected for the absorption is $4\times10^{40}$ erg s$^{-1}$.
The best fitted EW of the Fe \Ka and Fe \Kb lines are 4.1 and 1.2 keV, respectively.

To estimate the significance of the detection of each line, we use the Monte-Carlo method
following \citet{Mar06}. We generate 1000 fake spectra using the exposures and responses of 
the \cha observations for a null hypothesis model including Vapec, a continuum, and 
only the Fe \Ka and \Kb lines. 
For simplicity, we adopt the powerlaw continuum model in Figure 2, instead of the 
continuum model of MYTorus, both of which are similar in $2-8$ keV band. 
We then calculate the improvement of 
$\chi^2$ over the null hypothesis model by adding an emission line around the energy 
of each detected line (within a range of $\pm0.1$ keV). The cumulative distribution
of the $\Delta\chi^2$ is created for each line. We find the distributions are similar for different 
lines except the one of Ni line for $\Delta\chi^2<5$. Therefore we adopt an 
$\Delta\chi^2$ distribution averaged over all lines, which is plotted
in Figure 4. The estimated significance of the detection of each line is 
is listed in the last column of Table 1.
We see that the Si line is detected at $\sim3\sigma$, and the other fluorescent 
lines have a significance between 2 and 2.5 $\sigma$, while the Cr line has a 
significance of $\sim1.5\sigma$.

The unfolded spectrum for \nus data is plotted in Figure 5. 
There are totally about 1200 photons in the \nus spectra and
only the $3-50$ keV energy range is used due to the low S/N of the data above 50 keV.
We see that below $\sim10$ keV the spectrum is dominated 
by the ULX indicated in Figure 1, while above that the nucleus 
dominates. Because the exposure of the \nus observation is short, there are no 
many photons above 20 keV and the intrinsic spectrum is not well constrained. 
Further deep \nus observations will help to determine the intrinsic spectrum. 

To estimate the Fe abundance, we replace the two Gaussian lines of Fe \Ka and Fe \Kb with
the component of Fe fluorescent lines of MYTorus model and allow its normalization to vary. 
We find that the normalization of the line component of MYTorus is about 2 times 
that of the continuum component. Nevertheless, it does not necessarily
mean Fe is overabundant. \citet{GB15} showed that the EW of the  
Fe \Ka line could be enhanced by dust due to the reduction of the reflected 
continuum intensity caused by the smaller backscattering opacity of dust. 
This dust effect needs to be investigated to obtain the Fe abundance.

\begin{figure}
\hspace*{-0.5cm}
\includegraphics[height=2.4in]{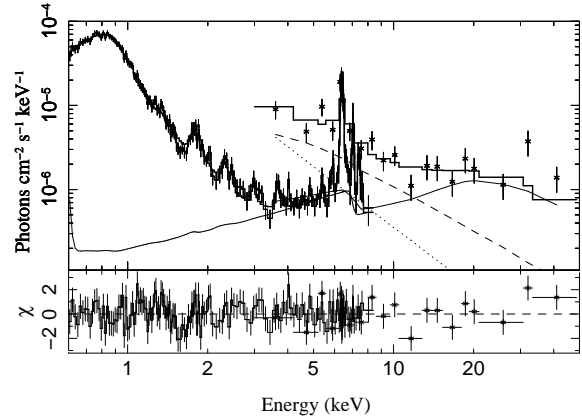}
\caption{
	The unfolded \nus spectrum of M51 (cross points) along with the fitted results (thick histogram).
	The continuum model of MYTorus (solid line) is over-plotted, together with the contaminating 
	powerlaw models of the ULX (dashed line) and the nuclear region excluding 
	the nucleus and the ULX (dotted line). Below $\sim10$ keV, the \nus spectrum is dominated 
	by the ULX, while above that it is dominated by the nucleus. 
	For comparison, the \cha spectrum and its unfolded MYTorus model are also plotted.
}
\end{figure}

\section{Discussion}

The Compton-thick nature of the nucleus of M51 is confirmed by the \nus observation 
of hard X-ray emission above 10 keV. The fitted equatorial column density of MYTorus
model is $\sim7\times10^{24}$ cm$^{-2}$. This value is consistent with the detection of 
strong HCN emission at the nucleus of M51, which suggests the presence of 
compact dense molecular gas \citep{Mat15}. A gas disk/torus with a density of $\sim10^6$ cm$^{-3}$ 
and a pc-scale size can provide the fitted column density.

We detected the faint fluorescent lines of neutral Si, S, Ar, Ca, Cr, 
and Mn, along with the prominent Fe K$\alpha$, Fe K$\beta$, and Ni \Ka lines,
from the deep \cha observation of M51. These lines can be used to measure their
abundances. As the reprocessing matter is Compton-thick, detailed radiative calculations,
such as those by \citet{MY09}, are needed to obtain accurate results.
Nevertheless, the relative abundance can be roughly estimated as follows.

The observed fluorescent lines are
most likely due to the scattering from a surface layer of unit optical depth that 
is not heavily absorbed by the torus itself.
The fluorescence flux of certain element $x$ can be expressed as follows
\citep[e.g.][]{GF91}

\begin{equation}
	{F_{x}}\propto Y_{x}\int_{E_x}^{\infty} AE^{-\Gamma}\frac{Z_x\sigma_{x,k}(E)}{\sigma_T(E)}\,dE,
\end{equation}
where $Y_x$ is the fluorescence yield, $AE^{-\Gamma}$ is the incident photon spectrum, 
$Z_x$ is the abundance relative to solar value, and
$\sigma_{x,k}$ and ${\sigma_T}$ are the 
absorption cross-sections of the K-shell electron of element $x$ and of all elements, respectively.
$\frac{Z_x\sigma_{x,k}}{\sigma_T}$ expresses the probability that a photon  
is absorbed by the K-shell electron of element $x$.
For $Z_{Fe}>1$, Fe dominates the opacity above the binding energy of Fe
and ${\sigma_T}$ is sensitive to $Z_{Fe}$.
This dependence is neglected for the rough estimate here. 
We also neglect the difference of the absorption of produced fluorescent photons.

The estimated abundances relative to Fe are listed in Table 1. The $\alpha$ elements of Si, S,
Ca, and Ar have an abundance pattern similar to that of the Sun. While for the trace element of
Mn, its abundance is about 10 times overabundant. This enhancement of Mn is likely due to 
the nuclear spallation of Fe caused by cosmic rays from the nucleus \citep{Ski97,TM10}.
In principle, it is possible that the soft Vapec component is due to the jet-heated gas,
which originally is part of the disk/torus gas emitting fluorescent lines. In this case,
the abundances of the Vapec model should be similar to those estimated from the fluorescent
lines. The relative abundance of Si, S, Ca, and Ar to Fe is about 1.5 for the Vapec model,
a little higher than those estimated from the fluorescent lines. This is expected since 
lower energy photons have a larger probability of being absorbed than higher energy photons.
Nevertheless, the details of the absorption depend on the distribution of the obscuring material 
and further modelling is needed for rigorous comparison.
These results show that the non-Fe fluorescent lines can be a valuable probe of 
the obscuring matter of AGN and the physical processes they related to.

As noted in \S 1, the EW of the Fe 6. 4 keV line of M51 is as large as 4 keV.
This large EW of the Fe line is due to the suppression of the illuminating intrinsic spectrum,
which also makes it possible to detect non-Fe fluorescent lines much fainter than the Fe line.
In fact, M51 is the only one with an EW of the Fe line larger than 2.5 keV in \citet{Lev02}.
It would be difficult to detect non-Fe fluorescent lines for samples with lower
EW of the Fe line. A search with other Compton-thick samples with an EW of the Fe line 
$\sim1-2$ keV is undertaking.

\section*{Acknowledgements}
We thank our referee for valuable comments.
JL is supported by NSFC grant 11203032, and LG is supported by 
the Strategic Priority Research Program ``The Emergence of Cosmological
Structures`` of CAS grant XDB09000000, NSFC grant 11333005, and NAOC grant Y234031001.
This research is based on data obtained from the \cha Data Archive and 
uses data from \nus mission.

\bibliographystyle{mn2e}

\begin{thebibliography}{}

\bibitem[Anders \& Grevesse(1989)]{AG89}
Anders, E. \& Grevesse, N. 1989, Geochimica et Cosmochimica Acta, 53, 197	

\bibitem[Comastri(2004)]{Com04}
Comastri, A. 2004, ASSL, 308, 245

\bibitem[Foster et al.(2012)]{Fos12}
 Foster, A. R., Ji, L., Smith, R. K. and Brickhouse, N. S. 2012, ApJ, 756, 128

\bibitem[Fukazawa et al.(2001)]{Fuk01}
Fukazawa, Y., Iyomoto, N., Kubota, A., Matsumoto, Y., Makishima, K. 2001, A\&A, 374, 73	

\bibitem[Fukazawa et al.(1994)]{Fuk94}
Fukazawa, Y., Makishima, K., Ebisawa, K., Fabian, A. C., Gendreau, K. C., Ikebe, Y., Iwasawa, K.,
Kii, T. et al. 1994, PASJ, 46, 141	

\bibitem[George \& Fabian(1991)]{GF91}
George, I. M. \& Fabian, A. C. 1991, MNRAS, 249, 352

\bibitem[Grimm et al.(2009)]{Gri09}
Grimm, H. J., McDowell, J., Fabbiano, G., Elvis, M. 2009, ApJ, 690, 128	

\bibitem[Gohil \& Ballantyne(2015)]{GB15}
Gohil, R. \& Ballantyne, D. R. 2015, MNRAS, 449, 1449	

\bibitem[Ho et al.(1997)]{Ho97}
Ho, L. C., Filippenko, A. V., Sargent, W. L. W. 1997, ApJS, 112, 315

\bibitem[Houck \& Denicola(2000)]{ISIS}
Houck, J. C. \& Denicola, L. A. 2000, in Astronomical Data Analysis Software 
and Systems IX, eds. Manset, N, Veillet, C., \& Crabtree, D. ASP Conf. Ser., 216, 591


\bibitem[Ikeda et al.(2009)]{Ike09}
Ikeda, S., Awaki, H., Terashima, Y. 2009, ApJ, 692, 608	

\bibitem[Kaastra \&Mewe(1993)]{KM93}
	Kaastra, J. S. \&  Mewe, R. 1993, A\&AS, 97, 443	

\bibitem[Liu \& Mao(2015)]{Liu15}
Liu, J. \& Mao, S. 2015, preprint (astro-ph/1505.07175)

\bibitem[Levenson et al.(2002)]{Lev02} 
Levenson, N. A., Krolik, J. H., Zycki, P. T., Heckman, T. M., Weaver, K. A., Awaki, H., Terashima, Y.
 2002, ApJL, 573, 81	

\bibitem[Markowitz et al.(2006)]{Mar06} 
Markowitz, A., Reeves, J. N., Braito, V. 2006, ApJ, 646, 783	

\bibitem[Matsushita et al.(2015)]{Mat15} 
Matsushita, S., Trung, D., Boone, F., Krips, M., Lim, J., Muller, S. 2015, ApJ, 799, 26	

\bibitem[Murphy \& Yaqoob(2009)]{MY09}
  Murphy, K. D. \& Yaqoob, T. 2009, MNRAS, 397, 1549	

\bibitem[Owen \& Warwick(2009)]{OW09}
Owen, R. A. \& Warwick, R. S. 2009, MNRAS, 394, 1741

\bibitem[Palmeri et a;.(2003)]{Pal03}
Palmeri, P., Mendoza, C., Kallman, T. R., Bautista, M. A., Melendez, M. 2003, A\&A, 410, 359	

\bibitem[Reynolds et al.(1994)]{Rey94}
Reynolds, C. S., Fabian, A. C., Makishima, K., Fukazawa, Y., Tamura, T. 1994, MNRAS, 268L, 55	

\bibitem[Shu et al.(2010)]{Shu10}
Shu, X. W., Yaqoob, T., Wang, J. X. 2010, ApJS, 187, 581	

\bibitem[Skibo(1997)]{Ski97}
Skibo, J. G. 1997, ApJ, 478, 522	

\bibitem[Stauffer(1982)]{Sta82}
Stauffer, J. R. 1982, ApJS, 50, 517

\bibitem[Tak$\acute{\rm a}$ts \& Vink$\acute{\rm o}$(2001)]{TV06}
Tak$\acute{\rm a}$ts, K. \& Vink$\acute{\rm o}$, J. 2006, MNRAS, 372, 1735

\bibitem[Terashima \& Wilson(2001)]{TW01}
Terashima, Y. \& Wilson, A. S. 2001, ApJ, 560, 139

\bibitem[Terashima \& Wilson(2004)]{TW04}
Terashima, Y. \& Wilson, A. S. 2004, ApJ, 601, 735

\bibitem[Turner \& Miller(2010)]{TM10}
Turner, T. J. \& Miller, L. 2010, ApJ, 709, 1230	

\bibitem[Yaqoob(2012)]{Yaq12}
Yaqoob, T. 2012, MNRAS, 423, 3360	

\end{thebibliography}

\end{document}